\documentclass[runningheads,citeauthoryear]{apinv39}
\usepackage{epsfig,cite,graphics}
\usepackage{amsmath}
\usepackage{graphicx}
\usepackage{caption} 
\usepackage{marvosym} 
\usepackage[utf8]{inputenc}

\begin{document}

\title{BV photometric observations of the flickering of the dwarf nova RX~And }
\titlerunning{Optical flickering of the dwarf nova RX~And}
\author{R. K. Zamanov$^1$, L. Dankova$^1$,  M. Moyseev$^{1,2}$, M. Minev$^1$, K. A. Stoyanov$^1$, V. D. Ivanov$^3$ \\ }
\authorrunning{Zamanov, Dankova, Moyseev, Minev, Stoyanov \& Ivanov}
\tocauthor{R. K. Zamanov, L. Dankova, M. Moyseev, M. Minev, K. A. Stoyanov, V. D. Ivanov} 
\institute{$ \; $\\
      $^1$Institute of Astronomy and National Astronomical Observatory, Bulgarian Academy of Sciences,
           Tsarigradsko Shose 72, BG-1784 Sofia, Bulgaria \\ 
	   $^2$Department of Astronomy, Faculty of Physics, Sofia University "St. Kliment Ohridski", \\ 
	   5 James Bourchier blvd., 1164 Sofia, Bulgaria  \\ 
	   $^3$European Southern Observatory, Karl Schwarzschild-Str. 2, 
	   D-85748 Garching bei Munchen, Germany
	   $ \; $ \\
	\email{rkz@astro.bas.bg,  lubav@astro.bas.bg \hskip 0.3cm   } }
\papertype{Research report, Submitted on 7.02.2023; Accepted on 15.03.2023}	
\maketitle

\begin{abstract}
We report photometry of the intranight variability of the  dwarf nova 
RX~And in two bands ($B$ and $V$). 
The observations are carried out during three nights in November-December 2022
at the 50/70~cm Schmidt telescope of the Rozhen National Astronomical Observatory. 
The observations indicate that the amplitude of the flickering is about 0.5 mag in $B$ band
when the star is in faint state ($m_V \approx 13.5$), but it is considerably lower (less than 0.1 mag) when the star is bright ($m_V \approx 10.9$). 
The mass accretion rate in high state is estimated to be $1.2 \times 10^{-9}$~M$_\odot$~yr$^{-1}$.
Combining our data and GAIA distances we find for the mass donor
in RX~And spectral type  K6V-K7V.  \\
The data are available upon request from the authors.
\end{abstract}
\keywords{Stars: dwarf novae -- novae, cataclysmic variables -- stars: individual: RX And}

\section{Introduction}

RX~And belongs to the subclass Dwarf Novae of the Cataclysmic Variable stars. 
The Cataclysmic Variables are compact close binaries (with orbital periods typically 1-12 hours) 
consisting of an white dwarf primary and a red dwarf secondary 
[e.g. Warner (1995), Sion \& Godon(2022) and references therein].
In the most cases the secondary is on the main sequence, fills in its Roche lobe 
and its Roche lobe overflow supplies material for  accretion disc around the white dwarf. 
The dwarf novae exhibit 
recurrent outbursts with amplitude of 2 to 5 mag on  time-scale of weeks-months, 
caused by disc instability and increase in the mass accretion rate. 

Following the AAVSO light curve generator, during the last three 
years (2020 - 2022), RX And varies in the range $10.9 <  V  \le 14.3$ 
with many dwarf nova outbursts during this period (one outburst every $\sim 30$ days). 

In this work, we present  quasi-simultaneous 
observations of the intranight variability of RX~And
in the Johnson $B$ and $V$ bands.

\section{Observations}
\label{obs}
The new observations of RX And are performed with the 50/70 cm Schmidt telescope of the 
Rozhen National Astronomical Observatory repeating $B$ and $V$ filters 
during three nights in November-December 2022. 
The telescope is equipped with a CCD camera 4096{\small x}4096 pixels.  
To reduce the readout time we binning the detector with a factor of 2{\small x}2
and windowed it down to 512{\small x}512  pixels, yielding 16'{\small x}16' field of view centered at RX~And. 
As comparison stars  we used: 
TYC 2807-1285-1 (01$^h$03$^m$55$^s$, +41$^\circ$15'22'', $B$=11.151, $V$=10.568) 
and 
TYC 2803-1045-1 ($01^h05^m10^s \; +41^\circ 14' 07''$ $B$=11.938, $V$=11.431). 
The $BV$ magnitudes are taken from the APASS DR10 (Henden et al. 2012). 

\begin{figure}    
   \vspace{17.0cm}     
   \includegraphics{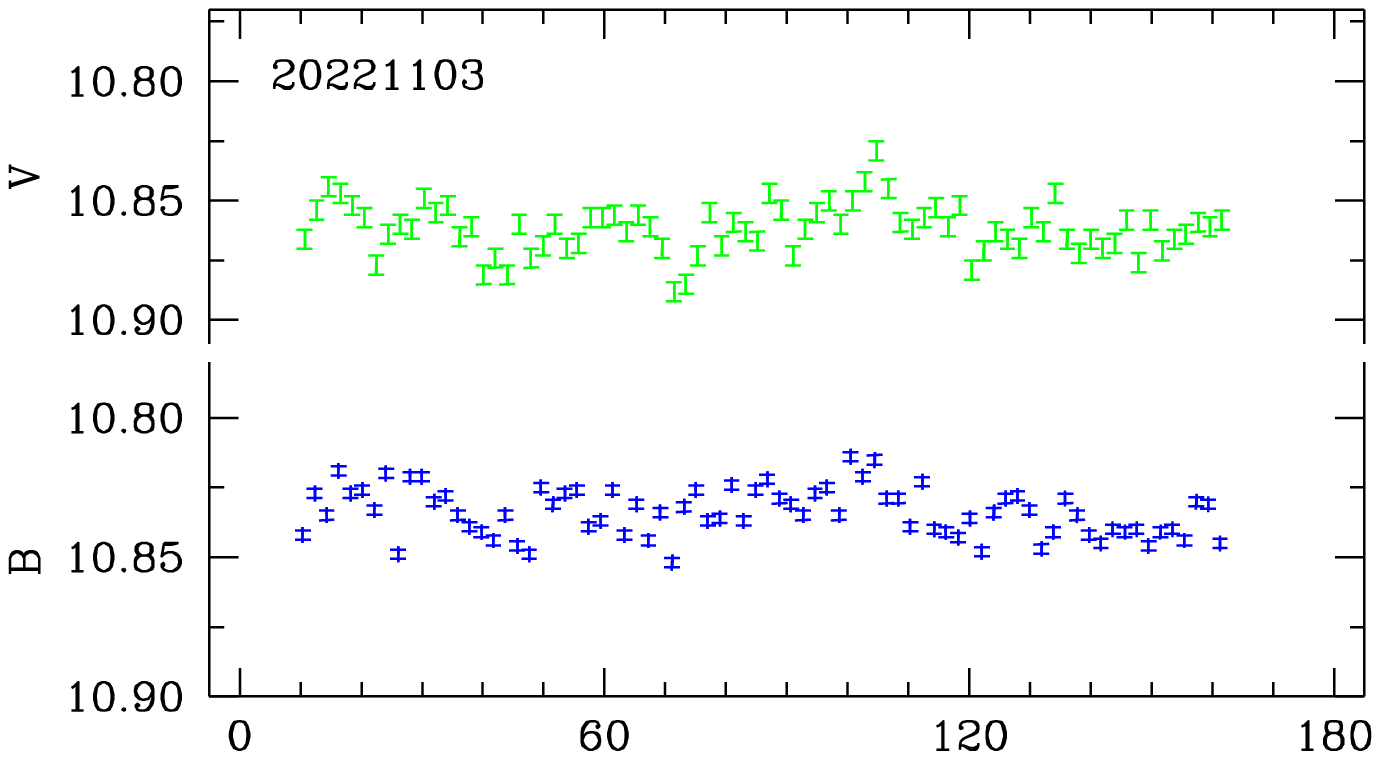}  
   \includegraphics{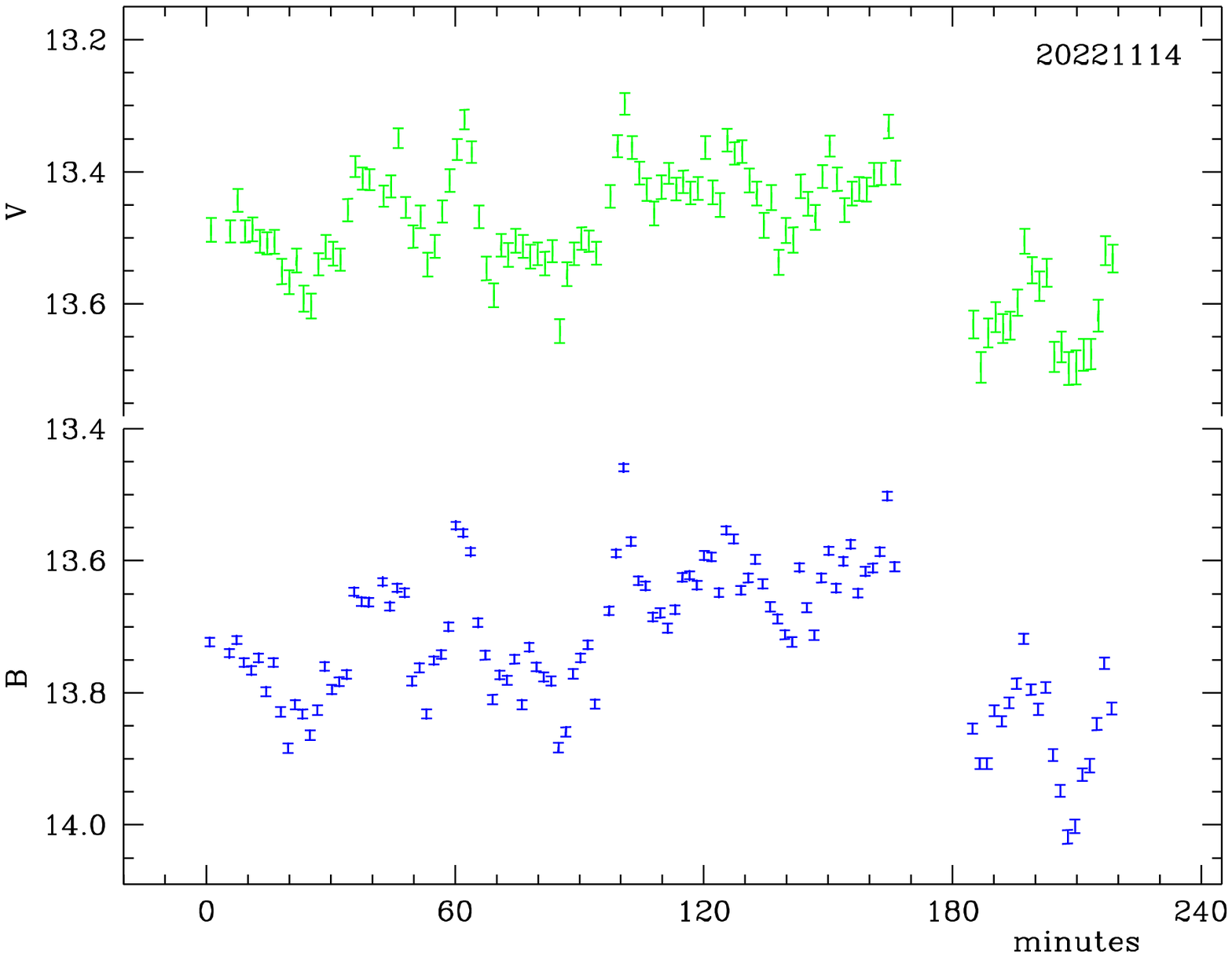}  
   \includegraphics{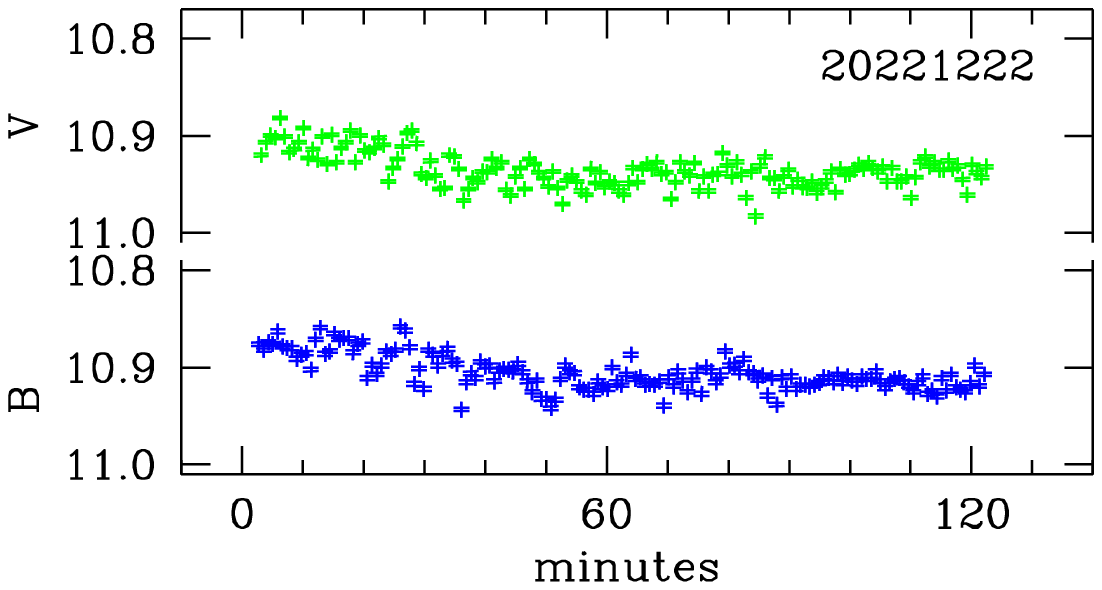}  
   \caption[]{Simultaneous observations 
   of the dwarf nova RX~And in $B$ and $V$ bands
   obtained on 3 November, 14 November, and 22 December 2022
   with the 50/70~cm Schmidt telescope of Rozhen NAO.  
   The X-axis is in minutes from start of the observation (1 hour = 60 min). 
   Note that the scaling of the axes are identical on all three panels. 
   The amplitude of the flickering 
   is higher when the star is fainter. }
   \label{f.RX} 
\end{figure}

\begin{figure}    
   \vspace{8.0cm}     
   \includegraphics{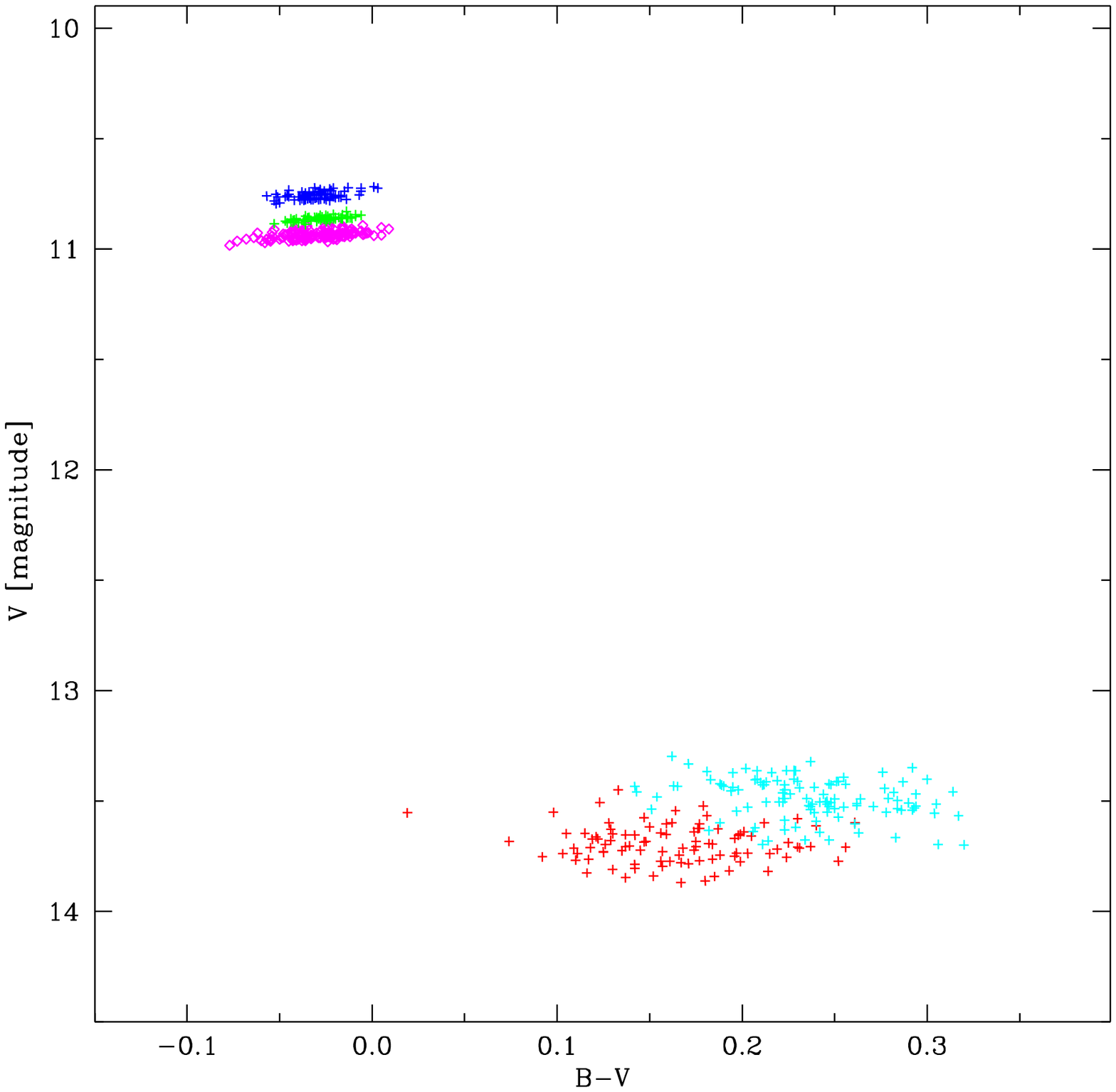}  
   \caption[]{Colour-magnitude diagram of RX~And. 
   Each night is plotted in different colour --
   20191025 (red), 20200102 (blue), 
   20221103 (green), 20221114 (cyan), 20221222 (magenta). 
   It is visible that the star becomes bluer as it gets brighter.
    }
   \label{f.cm} 
\end{figure}

The data reduction is carried out with IRAF (Tody 1993) 
following the standard recipes for processing of CCD images and aperture photometry. 
Our observations and results from the photometry are summarized in Table~\ref{t.obs}. We list the date and duration of the monitoring, the number of the exposures in each filter and the exposure times, the minimum (corresponding the maximum apparent brightness), maximum and average magnitude of the target, the standard deviation during the monitoring, the typical observational errors and the peak-to-peak variability magnitude. 
Our new observations are presented in Fig.~\ref{f.RX}.  
The colour-magnitude diagram (V versus B-V) of RX~And is presented 
in Fig.~\ref{f.cm}, where it is visible that the star becomes blues as it gets brighter. 
In the next section we will analyse the new as well
as the two runs 20191025 and 20200102 obtained with the same telescope setup and  presented in  
Zamanov, Nikolov \& Georgieva (2021). 



  \begin{table}[ht!]
  \begin{center}
  \caption{Log of RX~And observations and photometric results. Columns: date 
  (YYYY-MM-DD) and UT at start/end of the monitoring,
  filter, number of the frames and exposure times and light curve parameters 
  (see Sec.~\ref{obs}). 
  }
  \begin{tabular}{l l c l  r   c   | c   c c c  c  c c c c c}
  date  	  &  &  band & frames      &    &   &  min	 &  max   & average  & stdev  &  merr  & ampl.  \\
  UT start-end    &  &	&	      &    &   & [mag]   & [mag]  &  [mag]   &  [mag] &  [mag] & [mag]  \\
		  &	&	      &    &   &	&	 &	    &	     &        &        \\	 
 \hline
			 &	&	      &    &	&	 &        &         &        &	&	& \\	 
 
date 2022-11-03  &  &  B   & 78 x 10~s  &    &    & 10.814 & 10.852 & 10.8339 &  0.009 & 0.007    & 0.038 & \\
UT   20:43-23:17 &  &  V   & 78 x 7~s	&    &    & 10.829 & 10.888 & 10.8618 &  0.010 & 0.006    & 0.059 & \\
                 &  &	   &		&    &    &	   &	    &	      &        &	  &	  & \\    
date 2022-11-14  &  &  B   & 111 x 10~s &    &    & 13.459 & 14.018 & 13.7246 &  0.109 & 0.031    & 0.559 & \\
UT   21:10-24:51 &  &  V   & 111 x 7~s  &    &    & 13.297 & 13.698 & 13.4890 &  0.092 & 0.026    & 0.401 & \\
                 &  &	   &		&    &    &	   &	    &	      &        &	  &	  & \\
date 2022-12-12  &  & B    & 155 x 10~s &    &    & 10.858 & 10.943 & 10.9053 & 0.018  & 0.006    & 0.085 & \\
UT   18:46-20:46 &  & V    & 155 x 10~s &    &    & 10.882 & 10.983 & 10.9350 & 0.018  & 0.005    & 0.101 & \\
 \hline
 \label{t.obs}
 \end{tabular}
 \end{center}
 \end{table}
%

\section{Results and discussion}
\label{D.1} 

RX~And (2MASS J01043553+4117577) is a dwarf nova 
with orbital period $0.2098930$ days = 5h02m = 302.2 min  (Kaitchuck et al. 1988,  Kaitchuck 1989). 
The underlying white dwarf component of RX~And was first detected 
by Holm et al. (1991) using an International Ultraviolet Explorer (IUE) spectrum
during RX And's quiescence. The spectrum is obtained when the star was $V$=13.8 and
is consistent with the energy distribution of a 35000~K
hydrogen-atmosphere white dwarf.
The Hubble Space Telescope (HST) spectroscopy gives  
mass 1.14~M$_\odot$,  
temperature 34000~K, 
rotational velocity 600 km~s$^{-1}$ 
(Sepinsky et al. 2002, Sion et al. 2007,  Sion \& Godon 2012).

On the basis of  GAIA eDR3 (Gaia Collaboration et al. 2021), 
the model by Bailer-Jones et al. (2021) provides a distance to RX And, d = 196 pc.  
NASA-IPAC extinction calculator gives interstellar reddening $E_{B-V} < 0.06$ for  RX And. 
This upper limit refers to the extinction 
through the entire Milky Way in the direction of the object  
(in front of the object and behind it).
The calculator uses Galactic reddening maps to 
determine the total Galactic line-of-sight reddening, 
and is based on the results by  Schlegel, Finkbeiner \& Davis (1998)  
and Schlafly \& Finkbeiner (2011). 
Because the object is close to the Sun, likely there is no extinction, $E_{B-V} \approx 0$.

\subsection{Spectral type of the secondary}

The infrared observations show 
that many secondary stars in Cataclysmic variables have mid-K spectral type
(Howell 2005). For RX~And, Dhillon \&  Marsh (1995) found that the mass donor is a K5V star using 
infrared spectra. 
To verify this result with our data, we use
standard formula $M = m - 2.5 \log [ (d/10)^2 ]$,
and the minimum brightness in $V$ band.
In our data set the minimum brightness in $V$ band
is in the run 20191025, $V=13.871$ (see also Fig.~\ref{f.cm}). 
We find absolute $V$ magnitude of the secondary in the range  $7.4 <  M_V \le 8.2$.
The limits are calculated adopting fractional contribution of the secondary star to the total 
$V$ band flux in the range 100\% -- 50\%, respectively. It should be noted that the uncertainty in this fractional
contribution dominates the uncertainty in the spectral type, because it greatly exceeds the observational errors. 
The calibration of spectral types in absolute magnitudes 
(Sraizys \& Kuriliene 1981) gives $M_V=7.3$ for K5V star, and $M_V=8.1$ for K7V star. 
More recently, Pecaut \& Mamajek (2013)\footnote{The values are from version 2022.04.16 as given at\\
www.pas.rochester.edu/$\sim$emamajek/EEM\_dwarf\_UBVIJHK\_colors\_Teff.txt}  
lists similar values for the absolute magnites: $M_V=7.28$ for K5V star, $M_V=7.64$ for K6V, and  $M_V=8.16$ for K7V.

The range inferred from our observations indicates
that the secondary component in RX~And is most probably
of spectral type K6V - K7V, which is similar but a bit later than 
the spectral type given in Dhillon \&  Marsh (1995). 

\subsection{Amplitude of flickering}

Bruch (2021, 2022) noted that in some dwarf novae, the flickering amplitude is high during quiescence, drops quickly at an intermediate magnitude when the system enters into 
(or returns from) an outburst and, on average, remains constant above a given brightness threshold. 
Our observations of RX~And give a similar  result. 
In low state ($V \approx 13.5$), the peak-to-peak amplitude of the flickering in B band is
large: 0.56 mag (run 20221114) and 0.47 mag (run 20191025).
In high state ($V \approx 10.8$), the amplitude of the flickering in B band 
is small: 0.07 mag (run 20200102), 0.04 mag (run 20221103), 0.09 mag (run 20221212).

\subsection{Mass accretion rate}

For dwarf novae in high state, the main source of radiation in the optical bands is the accretion disc. 
In our observations (Fig.~\ref{f.cm}) the  highest brightness is $B$=10.687 and $V$=10.720 
in our run 20200201. The lowest brightness of RX~And is
$B$=14.040 and $V$=13.871 in our run 20191025.
We transformed these magnitudes into fluxes using the calibrations for 
a zero magnitude star (Rodrigo et al. 2012). 
We subtracted the minimum from the maximum flux, 
and the residuals should represent the contribution of the accretion disc.
In this way we find that the contribution of the 
accretion disc to the energy emitted in the optical B and V bands (in high state), 
is equivalent  to  $B=10.738$, $V=10.781$, and colour $(B-V)=-0.044$.  
For distance 196~pc, 
this corresponds to a black body with temperature $T=14220$~K, 
radius $0.305$~R$_\odot$, and luminosity 3.4~L$_\odot$.

The white dwarfs are objects in which the electron degeneracy pressure is equal to the gravitational pressure, and a remarkable property is that the more massive white dwarfs are smaller. 
To estimate the radius of the white dwarf in RX~And, 
we use the Eggleton's formula as given in Verbunt \& Rappaport (1988): 
\begin{multline}
\frac{ R_{wd}}{ R_\odot } = 0.0114 
\left[  \left( \frac{M_{wd}}{M_{Ch}} \right)^{-2/3} - \left( \frac{M_{wd}}{M_{Ch}} \right)^{2/3} \right]^{1/2}  \\
\times \left[  1 + 3.5 \left( \frac{M_{wd}}{M_p} \right)^{-2/3} + \left( \frac{M_{wd}}{M_p} \right)^{-1} \right]^{-2/3}, 
\end{multline}
where $M_{Ch}=$1.44~M$_\odot$ is the Chandrasekhar limit mass for the white dwarfs, 
$M_p $ is a constant $M_p= 0.00057$~M$_\odot$. 
We note in passing that the masses and radii estimated for
white dwarfs in detached eclipsing binaries agree with this formula
(e.g. Parsons et al. 2017).
%
%
This mass-radius relation gives radius $R_{wd} = 4370$~km for 
a $M_{wd} = 1.14$~M$_\odot$. 

The luminosity of the accretion disc is connected with the mass accretion rate:    
\begin{equation}
L_{disc} =  \frac{G \;  M_{wd} \;  \dot M_a}{2 \; R_{wd}},
\end{equation}
where $\dot M_a$ is the mass accretion rate, 
$M_{wd}$ is the mass of the white dwarf, 
$R_{wd}$ is the radius of the white dwarf. 
The underlying assumption in this equation is that the 
disc luminosity is half of the total accretion luminosity.
The other half is emitted in  UV/X-rays by the boundary layer between 
the  accretion disc and white dwarf
(more details can be found in Chapter 6 of Frank et al. 2012).
We find mass accretion rate for RX~And in high state 
$1.2 \times 10^{-9}$~M$_\odot$~yr$^{-1}$ ($7.5 \times 10^{16}$~g s$^{-1}$).
The error is about $\pm 20$\%, estimated from the inaccuracy of the parameters and the assumptions. 
This value is typical for the dwarf novae in high state and it is inside (in the upper part of)  the range $1 \times 10^{14} - 2 \times 10^{17}$~g s$^{-1}$
discussed in the theoretical models of dwarf novae accretion discs (e.g. Dubus et al. 2018).


\vskip 0.1cm 

\section{Conclusions}
We report  observations of the intranight 
variability of the dwarf novae RX And obtained in November-December 2022. 
The observations are performed quasi-simultaneously in two optical bands
($B$ and $V$) with the 50/70~cm Schmidt telescope of the Rozhen National Astronomical Observatory. 

We  find for the mass donor of RX~And 
absolute V band magnitude in the range  $7.4 < M_V < 8.2$ corresponding to spectral type  K6-K7V. 
We observed a large amplitude ($\approx$ 0.5 mag) flickering in low state. 
In high state the amplitude of the flickering is small $\approx 0.07$~mag. 
We  estimate that in high state, the mass accretion rate onto the white dwarf is
of about $1.2 \times 10^{-9}$~M$_\odot$~yr$^{-1}$.

\vskip 0.3cm 

{\small {\bf Acknowledgments: }
We acknowledge the partial support of this work by
the  Bulgarian National Science Fund (project  K$\Pi$-06-H28/2 08.12.2018
"Binary stars with compact object"). 
This research has made use of the SIMBAD database (operated at CDS, Strasbourg, France), 
the NASA/IPAC Extragalactic Database 
(operated by the Jet Propulsion Laboratory, California Institute of Technology,
under contract with the NASA)
and the AAVSO International Database contributed by observers worldwide. }

%


\begin{thebibliography}{}
\bibitem{} Bailer-Jones, C. A. L., Rybizki, J., Fouesneau, M., Demleitner, M., \& Andrae, R. 2021, AJ, 161, 147.
\bibitem{} Bruch, A., 2021, MNRAS, 503, 953. doi:10.1093/mnras/stab516 
\bibitem{} Bruch, A.\ 2022, MNRAS, 509, 4669. doi:10.1093/mnras/stab2675
\bibitem{} Dhillon, V.~S. \& Marsh, T.~R.\ 1995, MNRAS, 275, 89. doi:10.1093/mnras/275.1.89
\bibitem{} Drew, J.~E., Jones, D.~H.~P., \& Woods, J.~A.\ 1993, MNRAS, 260, 803. doi:10.1093/mnras/260.4.803
\bibitem{} Dubus, G., Otulakowska-Hypka, M., \& Lasota, J.-P.\ 2018, \aap, 617, A26. doi:10.1051/0004-6361/201833372 
\bibitem{} Frank, J., King, A., Reine, D., 2012, Accretion Power in Astrophysics, Cambridge University Press
\bibitem{} Gaia Collaboration, Brown, A. G. A., Vallenari, A. et al. 2021, A\&A, 649, A1. 
             doi:10.1051/0004-6361/202039657
\bibitem{} Henden A.~A., Levine S.~E., Terrell D., Smith T.~C., Welch D., 2012, JAVSO, 40, 430
\bibitem{} Holm, A.~V., Lanning, H., Mattei, J.~A., et al.\ 1991, JAAVSO, 20, 166 
\bibitem{} Howell, S.~B.\ 2005, ASP Conf. 330, 67
\bibitem{} Kaitchuck, R.~H., Mansperger, C.~S., \& Hantzios, P.~A.\ 1988, ApJ, 330, 305. doi:10.1086/166473
\bibitem{} Kaitchuck, R.~H.\ 1989, PASP, 101, 1129. doi:10.1086/132587 
\bibitem{} Parsons, S.~G., Gansicke, B.~T., Marsh, T.~R., et al.\ 2017, MNRAS, 470, 4473. doi:10.1093/mnras/stx1522
\bibitem{} Pecaut, M.~J. \& Mamajek, E.~E.\ 2013, ApJS, 208, 9. doi:10.1088/0067-0049/208/1/9
\bibitem{} Rodrigo, C., Solano, E., \& Bayo, A.\ 2012, 
                IVOA Working Draft 15 October 2012. doi:10.5479/ADS/bib/2012ivoa.rept.1015R
\bibitem{} Schlafly, E. F.,  Finkbeiner, D. P., 2011, ApJ, 737, 103  
\bibitem{} Schlegel, D. J, Finkbeiner, D. P., Davis,M., 1998, ApJ, 500, 525 
\bibitem{} Sepinsky, J.~F., Sion, E.~M., Szkody, P., et al.\ 2002, ApJ, 574, 937. doi:10.1086/341009
\bibitem{} Sion, E.~M., Godon, P., \& Szkody, P.\ 2007, ASP Conf. 362, 175
\bibitem{} Sion, E.~M. \& Godon, P.\ 2012, Memorie della Societa Astronomica Italiana, 83, 539
\bibitem{} Sion, E.~M. \& Godon, P.\ 2022, Galaxies, 10, 43. doi:10.3390/galaxies10020043  
\bibitem{} Straizys, V. \& Kuriliene, G.\ 1981, Ap\&SS, 80, 353
\bibitem{} Tody, D. 1993, ASP Conf., 52, 173 
\bibitem{} Warner B., 1995, Cataclysmic Variable Stars, Cambridge University Press, Cambridge
\bibitem{} Verbunt, F. \& Rappaport, S.\ 1988, Astrophysical Journal, 332, 193. doi:10.1086/166645
\bibitem{} Zamanov, R.~K., Nikolov, G., \& Georgieva, A.~T.\ 2021, 
             Bulgarian Astronomical Journal, 35, 110. doi:10.48550/arXiv.2104.12439 



\end{thebibliography}
\end{document}